\documentclass[12pt,preprint]{aastex}
\usepackage{amssymb}

\shortauthors{Wiebe \& Watson}
\shorttitle{Irregular Magnetic Fields and Dust Polarimetry}

\begin{document}

\title{Irregular Magnetic Fields and the Far-Infrared Polarimetry of Dust Emission
from Interstellar Clouds}
\author{Dmitri S. Wiebe\altaffilmark{1} and William D. Watson}
\affil{Department of Physics, University of Illinois, 1110 W. Green Street,
Urbana, IL 61801}
\email{dwiebe@inasan.ru; w-watson@uiuc.edu} 
\altaffiltext{1}{Permanent address: Institute of Astronomy of the RAS,
48, Pyatnitskaya str., 119017 Moscow, Russia}

\begin{abstract}
The polarized thermal radiation at far infrared and submillimeter wavelengths 
from dust grains in interstellar clouds 
with irregular magnetic fields is simulated. The goal is to determine how much
irregularity in the magnetic fields can be consistent with the observations that 
the maps of the polarization vectors are relatively
ordered.
Detailed calculations are performed
for the reduction in the fractional polarization and for the dispersion in position 
angles as a
function of the ratio of the irregular to the uniform magnetic field and as a 
function of the
relevant dimensions measured in correlation lengths of the field.
We show that the polarization properties of quiescent clouds
and of star-forming regions are consistent with 
Kolmogorov-like turbulent magnetic fields that are comparable in 
magnitude to the uniform component of the magnetic fields. If 
the beam size is much smaller
than the correlation length $L_{\rm corr}$ of the fields, the calculated 
percentage polarization
$p$ decreases to an asymptotic value when the number of correlation
lengths $N_{\rm corr}$ through the cloud exceeds a few tens.
For these values of $N_{\rm corr}$, the dispersion in the position 
angles $\sigma_\alpha$
is still appreciable---decreasing to only about $20^{\circ}$. However, when the finite 
size of a telescope beam
is taken into account, the asymptotic value of $p$ is reached for fewer 
correlation lengths (smaller 
$N_{\rm corr}$) due
to averaging over the beam; $\sigma_\alpha$ becames much smaller and consistent
with the observational data.
The smoothing of the polarization properties due to the combined effect of  
the thickness of the cloud and the finite size of the beam can be
described by a single variable which we designate as the generalized number of
correlation lengths.

In addition, we study various factors
that may contribute to the decrease in the linear polarization percentage with 
increasing intensity that is observed at submillimeter and far infrared wavelengths
in many, though not in all, dark clouds (the ``polarization hole" effect).
Depolarization due to a density
cutoff in the polarizing effect of dust, due to thermalization, and
due to correlations between the density and the properties of the 
magnetic field are considered.

\end{abstract}


\section{Introduction}
Since its
discovery in late 1940's (e. g., Hiltner 1956), 
the polarization of starlight through selective absorption by interstellar 
dust grains has been considered as a
valuable tool to study both the large-scale and small-scale structure of
interstellar magnetic fields. However, about ten years ago it was realized
that the polarization of starlight may not be a good probe of the magnetic
fields in dense interstellar clouds.

The polarization is caused by aligned grains and it is natural
to expect that the fractional polarization increases
when the light from a background star passes through a
region of higher extinction. However, the
observations give a different picture. Neither the percentage polarization, nor
the pattern of the polarization seem to differ appreciably for rays passing
through dense clouds in comparison with the rays that pass through the neighboring 
intercloud medium
\citep{mg91}. \cite{a98} demonstrated that the empirical relationship between
the maximum percentage polarization of starlight $p_{\max}$ and the optical
extinction $A_{\mathrm{V}}$ 
\begin{equation}
p_{\max}(\%)=3A_{\mathrm{V}}
\end{equation}
is only valid for stars that lie behind the ``general'' interstellar medium
and does not hold for stars that are located behind dense clouds. They 
concluded that in a dense medium for which $A_{\mathrm{V}}>1-2$, the
polarizing efficiency of dust grains is significantly reduced due to poor
alignment or due to changing grain properties (e.g., spherization). That is, 
the polarization of starlight cannot be a probe of the magnetic field
structure in dense quiescent and star-forming regions. This would also explain
why there is little correlation between the starlight polarization pattern
and quiescent cloud structure \citep{mg91} which would be expected in the
medium dynamically controlled by the magnetic field.

However, the observed inefficiency of the polarization of starlight in dense clouds
is at variance with the growing body of observations of the polarized
thermal emission by dust (e.g. Hildebrand et al. 2000). Significantly, this emission
is observed not only in regions of ongoing star formation where the
conditions for alignment can be different from those in dark clouds, but also in dark
and apparently starless cores \citep{firstmap,l183,jones}.
According to \cite{firstmap}, the
polarization of the emission at 850~$\mu$m by cold ($\sim10$~K) dust reaches 10\%
at the L1544 and L183 cores.
These cores have no internal sources of radiation, and their external illumination
should be negligible as some molecules (that are easily destroyed by UV radiation)
are present even in the outermost parts of these cores, well outside of regions
where the polarized emission is observed \citep{dickens,tafalla}.
This implies that $A_{\mathrm{V}}>3$ even at the core
peripheries; the extinction is greater in the inner parts of the cores ($A_{\mathrm{V}}>3$). 
Hence,
dust apparently preserves its polarizing properties at much higher
mean extinctions then those probed by starlight polarization. On the
other hand, there is recent evidence that a very low-mass protostar
is present in the L1014 core that previously was believed to be
starless \citep{young}. This result implies that
faint radiation sources may be embedded in some other ``starless''
cores as well, though such sources are presently undetected.

In our previous paper \citep{abs} we performed calculations for the linear
polarization of starlight due to extinction by aligned dust grains when the
starlight traverses a medium with irregular magnetic fields. We found that
the polarization properties of the starlight---the average fractional
polarization and the dispersion in position angles---can be essentially
unchanged if the rms of the irregular component in the optically thick
medium is greater than the average magnetic field. Thus, the observed
tendency for the starlight polarization to remain nearly constant as the light
passes through a dense cloud is understood as due to the changing properties of the
magnetic field without the need for a decrease in the polarizing 
efficiency of dust. In
this paper we explore the implications of this suggestion for the 
polarized emission by dust, and in particular, how much irregularity in the 
magnetic fields is consistent with the relatively ordered 
patterns that are observed for the polarization directions of the emitted 
radiation.

The breadths of spectral lines from interstellar clouds show velocity 
dispersions that exceed the thermal velocities. This and other evidence 
indicates that turbulent or wave motions are pervasive in the interstellar 
gas (e.g., Elmegreen \& Scalo 2004 for a current review). 
From MHD considerations, it follows that irregularities in the magnetic fields
are expected to be associated with such motions of the gas (e.g., V\'azquez-Semadeni et al. 2000).

The possible origin of polarization holes is also investigated.
In general, a polarization hole represents a factor
of a few decrease in percentage polarization as the intensity increases by
an order of magnitude, observed in many, though not in all, dark clouds. Its
name stems from the fact that it is mainly observed in centrally peaked
sources, though the anticorrelation of the percentage polarization and the
intensity may be independent on the source geometry \citep{omc3}.
This effect is wide-spread but not ubiquitous. For example, it does not
show up in the
NGC~7538 region observed by \cite{momose} at 850~$\mu$m. \cite{akeson}
observed the young stellar object NGC~1333/IRAS~4A with high angular
resolution (5$^{\prime\prime}$) and found polarized (4\%) dust emission
arising from very dense gas ($n>10^{8} $~cm$^{-3}$). An even higher percentage
of polarization (10\%) is found in the very center of the CB68 globule by
Val\'ee, Bastien, \& Greaves (2000). The conspicuous example of the
ambiguity about the polarization hole is represented by the Kleinmann-Low Nebula in
the OMC-1, where the hole is present when observed with low angular resolution
at 100~$\mu$m \citep{omc1} and disappears at higher resolution and larger
wavelength \citep{rao}. All this diversity seems to suggest that the
polarization hole effect is not related to a single mechanism but instead 
results from a combination of several mechanisms. In addition to the poor alignment of
grains and the loss of their polarizing ability, other contributing
factors may include specific variations of the large-scale magnetic field
and its unresolved small-scale structure that may become more complicated in
dense (possibly collapsing) parts of the star-forming region.

We first examine in detail how the irregularities in the 
magnetic field within a cloud reduce the fractional polarization 
and cause dispersion in the position angles of this polarization. That is,
we calculate  the ``polarization reduction factor" $F$ and the rms of the
position angles $\sigma _{\alpha }$ as a function of the ratio of the 
irregular component to the uniform component of the magnetic field, the 
dimension
of the cloud along the line-of-sight $N_{\mathrm{corr}}$ measured in 
correlation lengths of 
the field, and the beamsize of a telescope measured in correlation lengths. 
The only expression in the literature for $F$ of which we are aware
\citep{ld85}  
is applicable only in the limit for asymptotically large $N_{\mathrm{corr}}$ 
and negligible beamsize (even with these restrictions, we find that it is 
inaccurate for certain ratios of the irregular to the uniform component of 
the magnetic field). An expression is available for $\sigma _{\alpha }$
as a function of $N_{\mathrm{corr}}$ \citep{mg91}, but only for negligible beamsize.
Heitsch et al. (2001) have calculated the dispersion in polarization
angles of dust emission caused by irregular magnetic fields for specific,
turbulent MHD models. Although beamsize effects were included there, the 
focus of that study (the Chandrasekhar-Fermi relationship) was somewhat 
different from our investigation.

The paper is organized as follows. In \S~2 we describe the methods that are
used to generate the turbulent magnetic fields as well as 
the methods to calculate the emission of the polarized radiation from the 
cloud. In \S~3, the polarization characteristics of the emergent radiation 
are calculated as a function of the relevant parameters for the case of 
constant density. Effects of the non-uniform density are outlined in \S~4
where the emphasis is on considerations that might contribute to the 
polarization hole effect. A concluding discussion is given in \S~5.

\section{Basic Methods}

We describe the magnetic field as a sum of uniform and turbulent components.
As in our previous investigations, representative turbulent magnetic fields
are created by statistical sampling of the Fourier components of a power
spectrum with a Kolmogorov-like (i.e., power law) form and with Gaussian
distributions for the amplitudes (Wallin, Watson, \& Wyld 1998, 1999;
Watson, Wiebe, \& Crutcher 2001). Such methods are standard (e.g., Dubinski,
Narayan, \& Phillips 1995). As in our previous investigations, we focus
on a power spectrum that is somewhat steeper than Kolmogorov (here,
wavenumber $k^{-8/3}$ instead of the Kolmogorov $k^{-5/3}$) based in part on
the values of the correlation lengths that we extract (see Watson et al.
2001) from the results of MHD simulations by others (Stone, Ostriker, \&
Gammie 1998). Two quantities characterize the turbulent magnetic fields
created in this way---the rms value of each of the spatial components of the
irregular magnetic field $B_{\mathrm{rms}}$ (assumed isotropic) and the
correlation length of these components. The structure function for the 
magnetic field differs by less 
than 10 percent from its asymptotic value at a separation of $2/k_{%
\mathrm{min}}$. Here, $k_{\mathrm{min}}$ is the cutoff wavenumber introduced
to prevent an increase in the spectrum at wavenumbers that are smaller than
those at which energy is believed to be injected into the gas (Wallin et al.
1998).   We thus adopt the convenient expression $2/k_{%
\mathrm{min}}$ as an excellent approximation for the correlation length. 
With the customary assumption that alignment of the dust grains is
independent of the strength of the magnetic field, the behavior of the
linear polarization depends on the ratios of the strengths of the random to
the uniform magnetic fields 
\begin{equation}
b_{i}=B_{\mathrm{rms}}/B_{\mathrm{avg}}^{i},
\end{equation}
where $B_{\mathrm{avg}}^{i}$ is the uniform field component in $i$th
direction, and $i$ represents the field component either perpendicular to 
the line of sight (${\bot}$) and hence in the plane of the sky, or it 
represents the field component that is parallel to the 
line of sight (${\Vert}$). When there is no uniform
field in a given direction, $b_{i}=\infty$. Note that the average of the 
turbulent component is zero.

The fractional linear polarization of an emergent ray of radiation can be
expressed as 
\begin{equation}
P=R{\frac{C_{\mathrm{pol}}}{C_{\bot }+C_{\Vert }}}{\frac{\sqrt{q^{2}+u^{2}}}{%
N_{\mathrm{d}}}}  \label{peq}
\end{equation}
when the optical depth and fractional polarization both are small
(e.g., Wardle \& K\"{o}nigl 1990). Here, $R$ is a factor that accounts for
the imperfect grain alignment, $N_{\mathrm{d}}$ is the column density of
the dust grains, $C_{\bot }$ and $C_{\Vert }$ are absorption cross sections
perpendicular and parallel to the grain symmetry axis, and the effective
polarization cross section $C_{\mathrm{pol}}$ is given by 
\begin{equation}
C_{\mathrm{pol}}=\left\{ 
\begin{array}{ll}
C_{\bot }-C_{\Vert } & (\mbox{oblate grains}) \\ 
\frac{1}{2}(C_{\Vert }-C_{\bot }) & (\mbox{prolate grains}).
\end{array}
\right.
\end{equation}
The relative Stokes parameters are defined as 
\begin{equation}
q=\int n_{\mathrm{d}}^{\mathrm{a}}\cos ^{2}\gamma ^{\prime }\cos 2\phi \,%
\mathrm{d}s  \label{q_eq}
\end{equation}
and 
\begin{equation}
u=\int n_{\mathrm{d}}^{\mathrm{a}}\cos ^{2}\gamma ^{\prime }\sin 2\phi \,%
\mathrm{d}s.  \label{u_eq}
\end{equation}
In these expressions, $n_{\mathrm{d}}^{\mathrm{a}}$ is the number density of
aligned dust grains, $\gamma^{\prime}$ is the inclination of the total magnetic
field to the plane of the sky at a given point of the ray, and $\phi $ is
the angle between the projection of the total magnetic field onto the plane
of the sky and a reference direction---chosen here to be the direction of
the projection of the uniform magnetic field onto the plane of the sky. If $%
\phi =0$, $\gamma ^{\prime }=\mathrm{const}$, and $n_{\mathrm{d}}^{\mathrm{a}%
}$ does not vary along the ray, eq.~(\ref{peq}) reduces to 
\begin{equation}
P=R{\frac{C_{\mathrm{pol}}}{C_{\bot }+C_{\Vert }}}\cos ^{2}\gamma ^{\prime }.
\label{peq2}
\end{equation}
Lee \& Draine (1985) expressed the polarization reduction factor as 
\begin{equation}
\Phi \equiv RF_{\mathrm{LD}}\cos ^{2}\gamma,  \label{leedraine}
\end{equation}
where 
\begin{equation}
P={\Phi{\frac{C_{\mathrm{pol}}}{C_{\bot }+C_{\Vert }}}}
\end{equation}
to account for the polarization reduction due to a turbulent component of
the magnetic field ($F_{\mathrm{LD}}$), due to the imperfect grain alignment ($R$%
), and due to the inclination of the uniform magnetic field to the plane of the
sky ($\cos ^{2}\gamma $). In eq.~(\ref{leedraine}) 
\begin{equation}
F_{\mathrm{LD}}=\frac{3}{2}\left( \langle \cos ^{2}\theta \rangle -\frac{1}{3%
}\right)
\end{equation}
where the statistical average is over angles $\theta $ between the local
magnetic field and the direction of the uniform field. However, the
reductions due to turbulence and the inclination of the uniform field cannot, in
general, be separated as in these expressions. The focus of our work is on
computing the general polarization reduction factor $F$ due to turbulence, defined here as 
\begin{equation}
F=\sqrt{q^{2}+u^{2}}/\mathrm{N}_{\mathrm{d}}.
\end{equation}
to replace $F_{\mathrm{LD}}\cos ^{2}\gamma $ in equation (8).

The numerical integration in equation (5) and (6) is performed along the $%
128^{2}$ straight-line paths for rays that emerge perpendicular to the
surface at the grid points of our ``computational cubes'' in which magnetic
fields are generated at the $128^{3}$ gridpoints. Results for $F$ for the $%
128^{2}$ rays are then averaged to obtain the mean value of $F$, its standard
deviation $\sigma _{F}$, and the standard deviation of the polarization
position angle $\sigma _{\alpha }$ that are appropriate when the beam size
can be treated as ``infinitesimal'' (the position angle is obtained from the
standard relationship $\alpha =0.5\tan ^{-1}(u/q)$. Alternatively, for finite
beam sizes, the $q$ and $u$ are averaged separately over an appropriate
number of rays before $F$ and the direction of the linear polarization are
calculated.

\section{The Polarization Reduction Factor $F$ due to Turbulence}

In this Section, we examine the influence of irregularities in
the magnetic field on the linear polarization of the thermal emission by
dust under idealized conditions. Only the turbulent magnetic field is
allowed to vary along the ray. The density (except for the polarizations
based on MHD simulations for the fields), the degree of alignment and the 
emission by individual grains is taken to be constant and the medium is
assumed to be optically thin.
This model obviously deviates from real clouds which are not uniform
and isothermal, especially in regions of ongoing star formation.
In real objects where there are temperature variations, dust 
emission cannot be
presented as emission by a single component (e.g., Vaillancourt 2002).
There is evidence that temperature variations in clouds can be modest
(e.g., Tafalla et al. 2004).
\cite{torques} have proposed that radiative torques contribute significantly
in grain alignment. This suggests that the polarizing power of dust grains
also varies in real molecular clouds, being favored in regions exposed to
the radiation field of embedded stars or to the interstellar radiation field.
However, in dark molecular cloud cores, such as those mentioned in the Introduction, 
this may not be so
important. More critical for us are possible changes of the magnetic
field parameters that are directly or indirectly related to variations in the density. 
These are
neglected in our idealizations. Some of these issues will be addressed to a limited degree
in Section~4.

\subsection{Infinitesimal Beam Size}

In this subsection, we consider an ``infinitesimal'' beam size, i.e., the
size of the beam is much smaller than the characteristic scale length of the
turbulence---the correlation length. We first treat the case where there is
no component of the magnetic field that is parallel to the line-of-sight
(LOS). The behaviors of the polarization reduction factor $F$ averaged over
the 128$^{2}$ rays and the dispersion $\sigma _{\alpha }$ of the
polarization directions of these rays are shown in Figure~\ref{main} as a
function of cloud thickness (in correlation lengths) 
for clouds with two values of $b_{\bot}$ ($b_{\Vert }=0$) for which we also have fields from MHD
simulations. The polarization is seen to decrease from that due to a
completely uniform field ($F=1$). As the number of correlation lengths through 
the cloud increases, the polarization approaches the asymptotic value 
\begin{equation}
F_{\infty }=\frac{3}{2}\left( \langle \cos ^{2}\theta \rangle -\frac{1}{3}%
\right) =F_{\mathrm{LD}}
\end{equation}
of Lee and Draine which depends only on the ratio of the random to the
average magnetic field (also noted by Novak et al. 1997).

The number of correlation lengths needed to reach this asymptotic value
varies with $b_{\bot}$. If the uniform field is stronger than the turbulent field
($b_{\bot}=0.6$, hereafter the strong field case), the computed $F$ reaches 
$F_{\infty}$ for a cloud thickness of about ten correlation lengths. For the weaker uniform
field ($b_{\bot}=1.5$ hereafter the weak field case), $N_{\mathrm{corr}}\sim20-30$ 
is needed before the
resultant polarization is close to $F_{\infty}$. Note that the $b_{\bot}=0.6$
has been designated as ``medium'' in \cite{abs}.

The corresponding dispersion in the position angle is shown in right panel
of Figure~\ref{main}. It also depends on the number of correlation lengths
that have been traversed and on the magnetic field ratio, as shown by Myers
\& Goodman (1991). However, their expression (the dotted curves in Figure~%
\ref{main}) gives results that are similar to the computed $\sigma _{\alpha
} $ only for uniform fields that are rather strong and is less accurate when
the uniform magnetic field is somewhat weaker than the turbulent magnetic
field.

To justify our procedure for creating the magnetic fields, we also
present polarizations and dispersions that are computed with magnetic
fields which are the result of numerical simulations by others for
compressible MHD turbulence (Stone, Ostriker, \& Gammie 1998). They are
shown with open and filled triangles in Figure~\ref{main}.
In these simulations, the correlation lengths and
the magnetic field ratios are the same as in the fields created by the
statistical samplings that are used for the other computations in this
Figure. An important difference is that in computing $F$ and $\sigma
_{\alpha }$, the variations in the density of the dust along the rays are
included when we use the results of the MHD simulations. The polarizations
obtained with the weaker uniform magnetic field are in better agreement
with those calculated with our statistically created fields than are those 
obtained with the stronger uniform field, where larger variations in
density along the ray tend to reduce the polarizing power of the medium.

The simplification that the uniform magnetic field is parallel to the plane
of the sky is relaxed in Figure~\ref{los}, and a non-zero line of sight (${\Vert} 
$-component) $B_{\mathrm{avg}}^{\Vert}$ is allowed. Specifically, we compute $F$
for the same two values of $b_{\bot}$ as in Figure~\ref{main} with a range of
finite values for $b_{\Vert }$. Note that $\sigma _{\alpha }$ does not depend on $%
b_{\Vert }$ so that the anglular dispersions $\sigma _{\alpha }$ for the
computations in Figure \ref{los} are given by the right-hand panel of Figure \ref{main} for
the same $b_{\bot}$.

The solid lines with filled and open circles correspond to
the two cases shown in Figure~\ref{main}, where the uniform magnetic field
along the line-of-sight is zero. When $B_{\mathrm{avg}}^{\Vert}\neq 0$,
the asymptotic value depends upon $b_{\Vert }$ as would
be expected from equation (\ref{leedraine}), 
\begin{equation}
F=\frac{3}{2}\left( \langle \cos ^{2}\theta \rangle -\frac{1}{3}\right) \cos
^{2}\gamma.
\label{bznotzero}
\end{equation}
However, as can be seen from Figure~\ref{f_vs_b} where asymptotic values
$F_{\infty }$ are plotted versus $b_{\bot}$ for some representative values of
$\cos ^{2}\gamma $, the dependence of the computed $F_{\infty }$ on $\theta $
and $\gamma $ does not, in general, separate cleanly as in equation (\ref{bznotzero}). In
fact, equation (\ref{leedraine}) only seems to express the dependence of $F_{\infty }$ on
the line-of-sight magnetic field when $b_{\bot}\lesssim 0.5.$ For $b_{\bot}\gtrsim
1$, $F_{\infty }$ in Figure 3 becomes essentially independent of the
line-of-sight magnetic field. In other words, if the turbulent magnetic field
is only a factor of 1.5 greater than the component of the uniform magnetic field in the plane of
the sky, $F$ does not depend on the LOS magnetic field component,
whether it is weak or strong. This means that our conclusions about the polarization properties
of the dust thermal emission are essentially insensitive to the LOS magnetic field component for
the weak field case ($b_{\bot}=1.5$) that is considered, provided $\gamma<70^{\circ}$.
The number of correlation lengths needed to reach an asymptotic value is aproximately the same
for any $b_{\Vert }$.

\subsection{Finite Beam Size}

To examine the influence of the finite size of a telescope beam, we smooth
the computed map for $b_{\Vert }=0$ to represent various angular resolutions. As
might be expected, the important parameter in this case is not the ratio of
a beam size to the extent of the map, but the ratio $S$ of the beam size to
the correlation length $L_{\mathrm{corr}}$. This is illustrated in Figure~%
\ref{beamrat}, where $F$ and $\sigma _{\alpha }$ are presented for three
cases in which the ratio of the beam size to $L_{\mathrm{corr}}$ is the same
($S=1.6$), though the dimensions of the computational cubes as measured in
terms of correlation lengths are different. The areas of the cube surface
that are viewed by the beams are adjusted appropriately. Note that $L_{%
\mathrm{corr}}$ is expressed in Figure 4 and elsewhere as a fraction of the
length of an edge of a computational cube.

In Figure~\ref{beam}, $F$ and $\sigma _{\alpha }$ are presented for
several representative beam sizes as measured in correlation lengths. When
the beam size is much smaller than $L_{\mathrm{corr}}$, $F$ and $\sigma
_{\alpha }$ are the same as for the case of the infinitesimal beam.
However, if the beam size is larger than the correlation length, the
requirement that there be many correlation lengths through the cloud in order
for $F$ to reach $F_{\infty }$ is relaxed. The observed polarization
can be close to its asymptotic value even if $N_{\mathrm{corr}}<10$.

From the point of view of obtaining inferences about the strength of the magnetic
field, this means that low-resolution (or smoothed high-resolution) mapping
is preferable. The procedure described by Novak et al. (1997) for the large
$N_{\mathrm{corr}}$ limit can then be applied to determine the ratio of the
irregular-to-average components of the magnetic field even if $N_{\mathrm{corr}}$ is
relatively small or unknown. This straightforward way to infer the
irregular-to-average magnetic field ratio is valid only if the average magnetic
field is not inclined significantly to the plane of the sky (that is, $b_{\Vert }$
must be small). A large $||$-component of the magnetic field greatly complicates
the interpretation. In fact, an inspection of Figures~\ref{los} and
\ref{beam} shows that the curve in Figure~\ref{los} for a very strong magnetic
field ($b_{\bot}=0.6$ and $b_{\Vert }=0.15$) looks quite similar to the dotted curve
in the upper left panel of Figure~\ref{beam} which corresponds to a uniform
magnetic field that is weaker than the turbulent magnetic field, but is
observed with a large beam.

The rms angle $\sigma _{\alpha }$ decreases with increasing beam size, and
even for $b_{\bot}=1.5$ becomes smaller than $10$ degrees for clouds that
are no thicker than about ten correlation lengths when $S\sim 3$ in
Figure~\ref{beam}. This may explain why the measured dispersion of the
position angle in some observations is comparable to the observational error
 \citep{firstmap}.

For a given value of $b_{\bot}$ $(b_{\Vert }=0)$, both $F$ and $\sigma _{\alpha }$
in Figure~\ref{beam} depend on $N_{\mathrm{corr}}$ and $S$. However, we find
that to a reasonable approximation these two quantities can be combined into
a single quantity $G_{\mathrm{corr}}$ that describes their effect and can be understood
as a generalized measure of the number of ``correlation cells'' that are traversed
by rays received by a telescope. We define $G_{\rm corr}$ as the square of the sum
of the square root of the thickness of the cloud and
the area of the telescope beam expressed in correlation lengths
\begin{equation}
G_{\mathrm{corr}}=\left(N_{\mathrm{corr}}^{1/2}+S\right)^2.
\label{gcorr}
\end{equation}
The validity of this expression is demonstrated in the right-hand-side of
Figure~\ref{beam}, where $F$ and $\sigma _{\alpha }$ are shown as functions
of $G_{\mathrm{corr}}$. Equation~(\ref{gcorr}) shows that if $N_{\mathrm{corr}}$
is comparable to $S$, changes in $S$ tend to be more important for $F$ and
$\sigma_{\alpha }$ than are changes in $N_{\mathrm{corr}}$. When the number of
correlation lengths along a ray is large,
$G_{\mathrm{corr}}\approx N_{\mathrm{corr}}$. A rigorous analysis may give a more conceptually
based
relationship for $G_{\mathrm{corr}}$. However, the expression~(\ref{gcorr}) serves well
for the illustrative purpose in this paper.

Although $F$ and $\sigma _{\alpha }$ each depend on the thickness of the 
cloud and upon the beam size (as measured by $S$), the plot of 
$F$ as a function of $\sigma _{\alpha }$ does not depend upon these two
quantities.  Curves representing $F$ versus $\sigma_{\alpha }$ are shown
in Figure~\ref{f_vs_sigmaa} for two values of the magnetic field ratio
$b_{\bot}$. We have performed computations with a number of choices for
$N_{\mathrm{corr}}$ and $S$ to verify that the relationship between $F$ and
$\sigma_{\alpha }$ is, in fact, independent of these quantities.

Also shown in Figure~\ref{f_vs_sigmaa} are observational data for a selection of
large molecular clouds and individual globules (both with and without embedded sources).
The data for OMC-1, M17, W51, and W3 are taken from \cite{kuiper} and \cite{w3}.
Data for NGC~7538 are taken from \cite{momose}. Data for OMC-3 are taken from
\cite{omc3}. For
each object, the average polarization $\bar{p}$ and the ``true'' dispersions
of $p$ and $\alpha $ are computed from 
\begin{equation}
\sigma _{p}^{2}=\mathrm{var}\,{p}-1/\sum (1/\sigma _{p,i}^{2})
\end{equation}
and 
\begin{equation}
\sigma _{\alpha }^{2}=\mathrm{var}\,{\alpha }-1/\sum (1/\sigma _{\alpha,i}^{2}).
\end{equation}
where $\sigma _{p,i}$ and $\sigma _{\alpha ,i}$ are the observational errors
that are provided by the observers. We exclude points with $p/\sigma _{p}<3$
as unreliable, as well as points that apparently (based on the position
angle histogram) belong to a component that is different from the ``main''
component. Data for CB~26, CB~54, and DC 253--1.6 are taken from
\cite{henn2001}. As no tables are given in the latter paper, all the needed
numbers are extracted from the text. Data for L1544, L43, and L183 are
based on the paper by \cite{l183} and kindly provided by its authors.

The wavelengths of the observations
are indicated in the caption. It is well known that polarization is
wavelength-dependent. Hence, the data points obtained at different $\lambda$ cannot,
in principle, be compared directly, unless they are reduced to a
single wavelength with some correction factors like those implied by
Figure~15 from Hildebrand et al. (2000). However, there may be considerable
spread in these factors from object to object \citep{vail}, so we choose
to make no wavelength correction of any kind. In the relevant range of $\lambda$
($100-850\,\mu$m) these factors are near unity.

To relate the observed percentage polarization to the $F$ that is computed,
it is necessary to know the polarization that is produced by the grains when
the magnetic field is completely uniform. Hildebrand \& Dragovan (1995)
estimated that the maximum polarization produced by the mixture of perfectly
aligned silicate and graphite grains ($R=F=1$) should be about 35\% at $%
\lambda =100$~$\mu $m for their particular grain shape. We assume that the reduction
from the theoretical maximum of 35\% to the maximum polarization that is
observed ($\sim15\%$) occurs entirely
because of imperfect grain alignment and there is no contribution due to any
irregularity of the magnetic field ($R\simeq 0.4$). Thus,
to compare the observational data with the computed $F$, we
adopt $F_{\rm obs}=p_{\mathrm{obs}}/(R\cdot35\%)$.

The data points in Figure~\ref{f_vs_sigmaa} lie mostly in the
region bounded by the two computed curves. This tends to indicate that the
uniform magnetic field in these objects
is comparable with the random field or does not exceed
it by much. Of course, there is considerable uncertainty in relating the
observations to the computations in Figure~\ref{f_vs_sigmaa} because of the
poor knowledge of $R$. Without information about $R$, values of 
$\bar{p}$ and $\sigma_{\alpha }$ can only be used to provide general
guidance about the ratio of the irregular to the uniform magnetic field.
With this limitation in mind, Figure~\ref{f_vs_sigmaa} demonstrates the 
key result---that the observed average polarization properties of
interstellar clouds are consistent with irregularities in the magnetic field
and a uniform magnetic field that is relatively
weak, as we favored in Wiebe \& Watson (2001).

A non-zero component of the uniform magnetic field that is parallel to the 
line of sight does not affect
$\sigma_{\alpha }$, but does tend to reduce $F$ and hence shifts the computed 
curves downward in the left hand panel of Figure~\ref{f_vs_sigmaa}. As long as this parallel 
component is similar to or less than the perpendicular component of the uniform
field, its effect (see Figure~\ref{los}) will be to shift the curve for $b_{\bot}=0.6$ 
downward by no
more than about 0.1 in $F$ and to leave the curve for $b_{\bot}=1.5$ essentially 
unchanged. The data in Figure~\ref{f_vs_sigmaa} would still be consistent with the relatively
weak uniform magnetic field.

A quantity that is free from the uncertainties in $R$ is
the relative polarization dispersion, $\sigma _{F}/F$.
In the right hand panel of Figure~\ref{f_vs_sigmaa}, this quantity is plotted as
a function of $\sigma_{\alpha}$. The relationship $\sigma_\alpha=28.6^{\circ}\sigma_F/F$,
that would be expected for statistically independent $q$ and $u$, only
holds for small $\sigma_\alpha$. For $\sigma_\alpha>10^{\circ}$,
the value of $\sigma _{F}/F$ is only weakly sensitive to $b_{\bot}$.
It also does not depend on the beam
size, though this is not shown explicitly. On the other hand, 
it grows somewhat for large $b_{\Vert }$. Thus, a relative
polarization error that is greater than unity can be indicative of a 
significant line-of-sight
component of the magnetic field. With the few exceptions,
almost all the observational points are concentrated near
the computed values. Note that $\sigma _{\rm{F}}$ and $\sigma _{\alpha }$ in
Figure~\ref{f_vs_sigmaa} are not observational errors, but actual
dispersions caused by the variations of the magnetic field.

In Figure~\ref{map} we show the maps of polarization vectors that have been
used to compute the average
values presented in the foregoing. Maps are shown for two values of $N_{\rm corr}$---the
number of correlation lengths traversed along each ray. The maps are smoothed with
beam sizes of $S=1.6$ and 3.2. They correspond to dotted and short-dashed lines on the
leftmost panels of Figure~\ref{beam}.  While the mean
polarization is close to its asymptotic value on these maps, there are regions where $F$
exceeds this value almost by a factor of 2. Even though the uniform magnetic field
is weak, the polarization shows a regular pattern at $N_{\rm corr}\sim30$,
when the beam size is comparable to the correlation length, and at $N_{\rm corr}\sim10$,
when the beam size is greater than $L_{\rm corr}$.

We conclude that the irregularities in the magnetic field used in Wiebe \& Watson (2001)
to understand the polarized absorption of starlight are not in conflict with the observed 
polarization properties of the thermal radiation that is emitted by dust grains,
provided that
the correlation lengths in the regions that have been investigated are somewhat smaller 
than the sizes of the telescope beams. It is more difficult to understand how a
polarization hole can appear in such an environment.
Along with the regions of high polarization, there are some spots in Figure~\ref{map}
where $F$ is much smaller than the mean value. Inspection of the underlying
map for the infinitesimal beam size shows that this is due to the higher than average
magnetic field tangling in the particular regions. If for some reason
disturbances in the magnetic fields in actual gas clouds are associated with the 
density enhancements, such regions
will be observed as polarization holes. This issue is further addressed in the next Section.
However, in general, it is unlikely that a polarization holes will be due to this cause alone.
From Figure~\ref{main}, the decrease in the percentage polarization
can be caused either by an increase in the number of correlation lengths or
by a weakening of the mean magnetic field (or both) in a dense region
where the hole is observed. However, in the weak uniform field case and at $N_{\rm corr}>10$,
which is preferable for our proposed interpretation of the polarization of starlight, 
$F$  already is about as small as it can be
(especially, when the finite beam size is taken into account), leaving little space
for further decrease.

On the other hand, the magnetic field structure may cause polarization holes
if for some reason the uniform field is stronger on a periphery of a dense object
and weakens closer to its center. If the
decrease of polarization with growing intensity is caused by
shorter correlation lengths, one would expect that $\sigma _{\alpha }$ is
smaller in dense cores than in the surrounding medium. Interesingly, this is
what actually is observed in some cores of Barnard 1 dark cloud \citep{MW2002}.

\section{Density Effects}

In the previous section we concluded that the unresolved magnetic field
structure cannot be a common reason for the polarization hole effect.
In this Section we investigate whether the polarization hole effect can 
be due to effects that are related to the variations in density.

We generate the spatial distribution of dust by the same procedure that 
we used to create the
magnetic field. A single component of the vector field is created in the
Fourier space with the same $k_{\min}$ as is used to generate the magnetic
field. This distribution of Fourier waves is inverted to find the
distribution in coordinate space. The resultant Gaussian quantity is
interpreted as the logarithm of the density. The density created in this way has a
log-normal probability distribution function (PDF) that possesses the
desired spatial variations as characterized by the correlation length. The
volume weighted distribution of the quantity $y\equiv\log(n_{\mathrm{d}}/%
\bar{n}_{\mathrm{d}})$, where $n_{\mathrm{d}}$ and $\bar{n}_{\mathrm{d}}$
are the actual and spatially averaged densities of dust grains, is 
\begin{equation}
f(y)={\frac{1}{\sqrt{2\pi\sigma^{2}}}}\exp\left[-{\frac{(y+|\mu|)^2}{2\sigma
^{2}}}\right].
\end{equation}
The log-normal PDF seems to be relevant in the isothermal case and is
reproduced in many MHD simulations (Nordlund \& Padoan 1999;
V\'{a}zquez-Semadeni et al. 2000; Ostriker et al. 2001). In Figure~\ref
{dustdis} we compare the PDFs for our simulated densities with that
of the MHD simulation by Stone et al. (1998). The
corresponding $|\mu|$ values are given in the legend. To assess the 
importance of the spatial
structure of the density field, distributions with different numbers of
correlation lengths ($N_{\mathrm{corr}}=5$ and 12) through the computational
cube are considered. The polarization reduction factor $F$ in this Section 
thus includes the reduction of
polarization due not only to tangling of the magnetic field, but also 
due to the various additional factors that are studied here.
The correlation length is assumed to be independent of density.

\subsection{A Density Cutoff for Polarization}

In this subsection we examine the consequences of the
assumption that dust grains lose their ability to
polarize light as the density increases. This might be due to poor 
alignment caused by more frequent collisions with the gas molecules
or to the growth of icy mantles that make
the grains more spherical. A similar suggestion was recently studied by Padoan et al.
(2001). They showed that it is possible to reproduce the observed
anticorrelation between $P$ and $I$ under the assumption that grains are not aligned
at depths beyond a certain value of extinction $A_{\mathrm{V}}$ measured from the 
edges of the clouds. This assumption
implies that radiative torques are the primary cause for grain alignment.

Here, we assume that grains are no longer aligned when the density exceeds a 
certain critical density $n_{\lim}$. That is, in
equations~(\ref{q_eq}) and (\ref{u_eq}) we assume that 
\begin{equation}
n^{\mathrm{a}}_{\mathrm{d}}=\left\{ 
\begin{array}{ll}
n_{\mathrm{d}} & n_{\mathrm{d}}\le n_{\lim} \\ 
0 & n_{\mathrm{d}}>n_{\lim}
\end{array}
\right..
\end{equation}

First we consider the density distribution that corresponds most closely to 
the distribution from the
MHD calculations (labeled `N' in Figure~\ref{dustdis}). Intensity maps 
overlaid with the
polarization vectors and the $F$ vs $I$ scatter diagrams for the `N'
distribution are shown in Figures~\ref
{narrowmed} and \ref{narrowweak}. The density cutoff is set as 1.7 times 
the median density. The upper
panels of the Figures correspond to calculations for which the number of 
correlation lengths along
the side of the map $N_{\mathrm{corr}}\approx5$. In the calculations for 
the lower panels $N_{\mathrm{%
corr}}\approx12$. The uniform magnetic field is strong ($b_{\bot}=0.6$) in 
Figure~\ref{narrowmed}, and is weak ($b_{\bot}=1.5$) in Figure~\ref{narrowweak}. 
It is noteworthy that the densities at almost
30\% of the grid points exceed the specified critical density and thus
do not contribute to $Q$ and $U$. If we assume that the side of the
computational cube is 1~pc (so that the diameter of the central
concentration in upper panels is about a few tenths of a parsec
and $L_{\rm corr}\sim0.1$~pc), the 
optical depth $\tau\approx10^{-3}$ for the
`N' distribution corresponds to a
median gas density of approximately $10^{4}$~cm$^{-3}$ at $\lambda=1.3$~mm
(where a dust opacity $\kappa _{1.3\,\mathrm{mm}}=1$~cm$^2$~g$^{-1}$ is assumed;
Ossenkopf \& Henning 1994). To produce the observed
decrease in polarization, dust grains should be unaligned in all clumps 
that are at least moderately dense for these calculation with the `N' 
distribution. This seems to contradict the fact that highly
polarized dust emission is sometimes observed from regions that are even more dense.

Even under these rather extreme conditions, the 
scatter diagram for $F$ vs 
$I$ for the weak field case is far from the tight appearance actually seen 
in many
observations---at least, at long wavelengths. In the strong field case, the
tightness of the correlation depends upon the specific realization at this 
$n_{\lim}$; i.
e. in some statistical realizations of the density the scatter can be 
more significant.
If $n_{\lim}$ is increased by only by a factor of a few, the correlation
between the percentage polarization and the intensity completely disappears.
Lower values of $n_{\lim}$ tend to produce tight correlations in any
realization, though the percentage polarization is quite low. If
we set $n_{\lim}$ equal to the median density, the maximum $F$ (at the
lowest intensity) is only about 0.1 even for the strong field. The observed
slope---a decrease by a factor of about 3 in percentage polarization 
as the intensity
grows by an order of magnitude---is only reproduced by the strong field
and low $N_{\mathrm{corr}}$.

In other models of HD and MHD turbulence the density distribution can be
wider, with the range in densities reaching six orders of magnitude (Padoan et
al. 1998; Klessen, Heitsch, \& Mac Low 2000). Hence, we
introduce a wider PDF labeled `W' in Figure~\ref{dustdis}. With the same
choice of fiducial parameters, the median density is $\sim4\times10^{3}$ cm$%
^{-3}$. The high density tail of the distribution extends to $\sim10^{8}$~cm$%
^{-3}$, albeit the number of high density cells is small. Thus, only a tiny
fraction of the volume contributes to the dust emission. In such a wide
distribution, the tight correlation between $F$ and $I$ can be reproduced
with a much higher value of $n_{\lim}$. Maps and diagrams for the wide density 
PDF and $%
n_{\lim}\approx50n_{\mathrm{med}}$ are shown in Figures~\ref{widemed} and \ref
{wideweak}. Only 4\% of the grid points do not contribute to the polarized 
emission by this criterion. This
seems sufficient to obtain (at least in most realizations) a rather tight 
$F$ and $I$ correlation in the strong field case. The weak magnetic field
tends to produce more scatter, though $F$ at low
intensity is quite high in all four cases that we present--- reaching value 
of 0.7 for the strong uniform field and $N_{\mathrm{corr}}=5$. In all cases 
except the last, $F$
decreases significantly with intensity, in agreement with
observations. To summarize, increasing $n_{\lim}$ leads to more scatter; 
decreasing 
$n_{\lim}$ produces tighter correlations but smaller $F$.

\subsection{Thermalization Effects}

It is widely assumed that the emission by dust from the star forming regions is
optically thin. Nevertheless,  unresolved optically thick
clumps are sometimes mentioned (e.g. Gull et al. 1978; Schleuning 1998;
Schleuning et al. 2000) along with other possible causes for the decrease 
in the percentage polarization in the brightest regions of dense clouds. 
Clearly, this mechanism is most likely to be effective in star
forming regions where high opacities and submillimeter luminosities are expected
for Class~0 objects (Andr\'e, Ward-Thompson, \& Barsony 2000). Optical
depths of order of~1 are implied both at the far-infrared (Larsson et al.
2000; Mookerjea et al. 2000) and in the submillimeter range (Sandell 2000; Sandell
\& Knee 2001) in the dense parts (cores) of at least some star forming regions.

We will examine, to a limited degree and only in Figure~\ref
{thermal}, the consequences of the thermalization that occurs
when the optical depth approaches and
exceeds one. For this, the previous expressions must be generalized
somewhat. The optical depth becomes
\begin{equation}
\tau=\tau_{0}\int n_{\mathrm{d}}ds
\end{equation}
where $\tau_{0}$ is the normalization factor, and the expressions for relative
Stokes parameters are 
\begin{equation}
\begin{array}{l}
i=1-e^{-\tau} \\[0.1cm] 
q=\tau_0e^{-\tau}\int n_{\mathrm{d}}\cos^{2}\gamma^\prime\cos2\phi ds \\[0.1cm] 
u=\tau_0e^{-\tau}\int n_{\mathrm{d}}\cos^{2}\gamma^\prime\sin2\phi ds.
\end{array}
\end{equation}
These expressions allow the emission by dust to be optically thick, but
still assume that the fractional polarization is small.

To investigate the effect of large optical depth, we utilize an even wider
density distribution (labeled `T' in Figure~\ref{dustdis}) to calculate
$F$ for three different values of $\tau_{0}$ which are chosen so that 
the maximum optical depth along any ray is $\tau_{\max}=1$,
or 10, or 100. There is no density cutoff for grain alignment in this case. 
Results for the second
and the third cases are shown in Figure~\ref{thermal}. At $\tau_{\max}=1$,
no trend of decreasing polarization with increasing intensity is evident. 
It starts to appear at $\tau_{\max}=10$, where the properties of
the density field are still realistic. With the adopted fiducial parameters,
at $\lambda=1.3$~mm the `T' density PDF results in a few rarified cells ($%
n\sim1$~cm$^{-3}$) and a few very dense cells (with the peak gas density $%
n\approx2\times10^{10}$~cm$^{-3}$). The median density for the entire cube is $%
10^3$~cm$^{-3}$, and median optical depth is $\sim5\times10^{-3}$ (for the
far-infrared waveband, all densities are an order of magnitude lower due to
the higher opacity).

However, a slope that is similar to what is observed is only achieved when 
$\tau_{\max}=100$ and with a
median optical depth $\sim10^{-2}$, which probably is too high for the
submillimeter waveband. Thus, we conclude that thermalization probably is
not important for explaining the polarization holes observed at longer
wavelength, though it may contribute to this effect in the far-infrared.

\subsection{Effects of Correlations between the Density and the Magnetic Field}

The procedure that we use to generate the matter densities does 
not lead to distributions
for the matter and magnetic fields that are consistent with one another. 
Specifically, in the previous subsections we used the same ratio of the turbulent
to the uniform magnetic field $b_{\bot}$ for the entire cube, despite the
variations in density. In reality, we may expect that the spatial distributions 
of density and of the magnetic field 
are not independent. In Figure~\ref{mfcolden} we plot the standard
deviations of the 3D~magnetic field along each ray versus the column density
in arbitrary units for an MHD simulation (weak uniform field case) by Stone
et al. (1998). Though the
lower part of the plot is more densely populated, a trend is evident. 
For clarity, only each 100th
point is shown for column densities less than 400. A similar trend is
observed in the cube computed with the strong uniform field. Thus, in the available
MHD simulations, the magnetic field does exhibit a stronger variation along the
rays with higher optical depth--- though the range in density is not wide.

In our model, the ability to reproduce this 
correlation between the matter density and the magnetic field is quite limited.
To evalute its importance semi-quantitatively, we calculate a model in
which $b_{\bot}$ is proportional to the local density and is normalized to be
equal to a specific value at the median density, designated as $b_{\mathrm{med}}$. 
The `N' density PDF is
utilized at $N_{\mathrm{corr}}=12$ (these parameters are chosen to reproduce
the results of the MHD simulations as closely as possible). There is no density cutoff
for the alignment of the grains. In Figure~\ref{bd} we show $F$ and $I$ for the calculations 
with $b_{%
\mathrm{med}}=1.6$ and $b_{\mathrm{med}}=0.5$. For the two cases that are presented, 
the magnetic field ratios in
most grid cells correspond to those of our weak and
strong fields, respectively. At the low density end of the distribution,
the magnetic field is strictly uniform; at the high density tail, there is
essentially no uniform field.

The decrease in $F$ with increasing intensity is present in both cases, 
though only the strong field ($b_{\rm med}=0.6$) reproduces
the observed slope. If we use the same scaling of $b_{\bot}$ with density and the
`W'~density PDF, the trend is absent both in the weak and the strong field
cases. It only reappears at $b_{\mathrm{med}}<0.2$. However, this scaling 
probably is not appropriate for the `W' distribution, which is intended to
describe the more advanced evolutionary stages of the star forming regions.

\section{Discussion}

The polarization of starlight due to absorption by dust and the polarization 
of the thermal emission
from dust provide two ways to study the structure of the magnetic field in star
forming regions. However, observations indicate that tracing 
the direction of the magnetic field with polarized 
starlight can be seriously
hindered by some factor or factors that prevent the grains in the dark interiors of
clouds from influencing the observed polarization. In \cite{abs} we argued
that a relatively strong turbulent magnetic field within dark clouds
may sufficiently randomize the orientations of the dust that their contributions
to the polarization of starlight effectively cancel.
If a ray of starlight is initially polarized by 
passing through the general interstellar medium with $%
b_{\bot}\sim0.5$, and then traverses about 10 correlation lengths within a cloud
where $b_{\bot}$ is about 3 times greater, its polarization is
almost unchanged in magnitude and in direction as a result of passing through 
the cloud.

In this paper we show that the occurence of such irregular magnetic
fields in dark clouds is not in conflict with the relatively ordered polarization
vectors of the dust emission observed both in quiescent dense cores and in
the regions of ongoing star formation. The  maps of polarization
vectors that are computed using a medium with relatively strong turbulent magnetic 
fields and with large
$N_{\rm corr}$ show the regular patterns typically seen in the infrared and
submillimeter observations. The average percentage polarization within these
maps is $2-4\%$ (assuming $R\approx0.4$), and is close to what is observed.
Heitsch et al. (2001) reached similar conclusions about the polarized emission
for certain specific turbulent MHD models for the medium.
Chance misalignment of polarization vectors in adjacent points can cause
a decrease in the percentage polarization---reminiscent of polarization
hole---if this region is observed with low angular resolution.

\begin{figure}[t]
\plotone{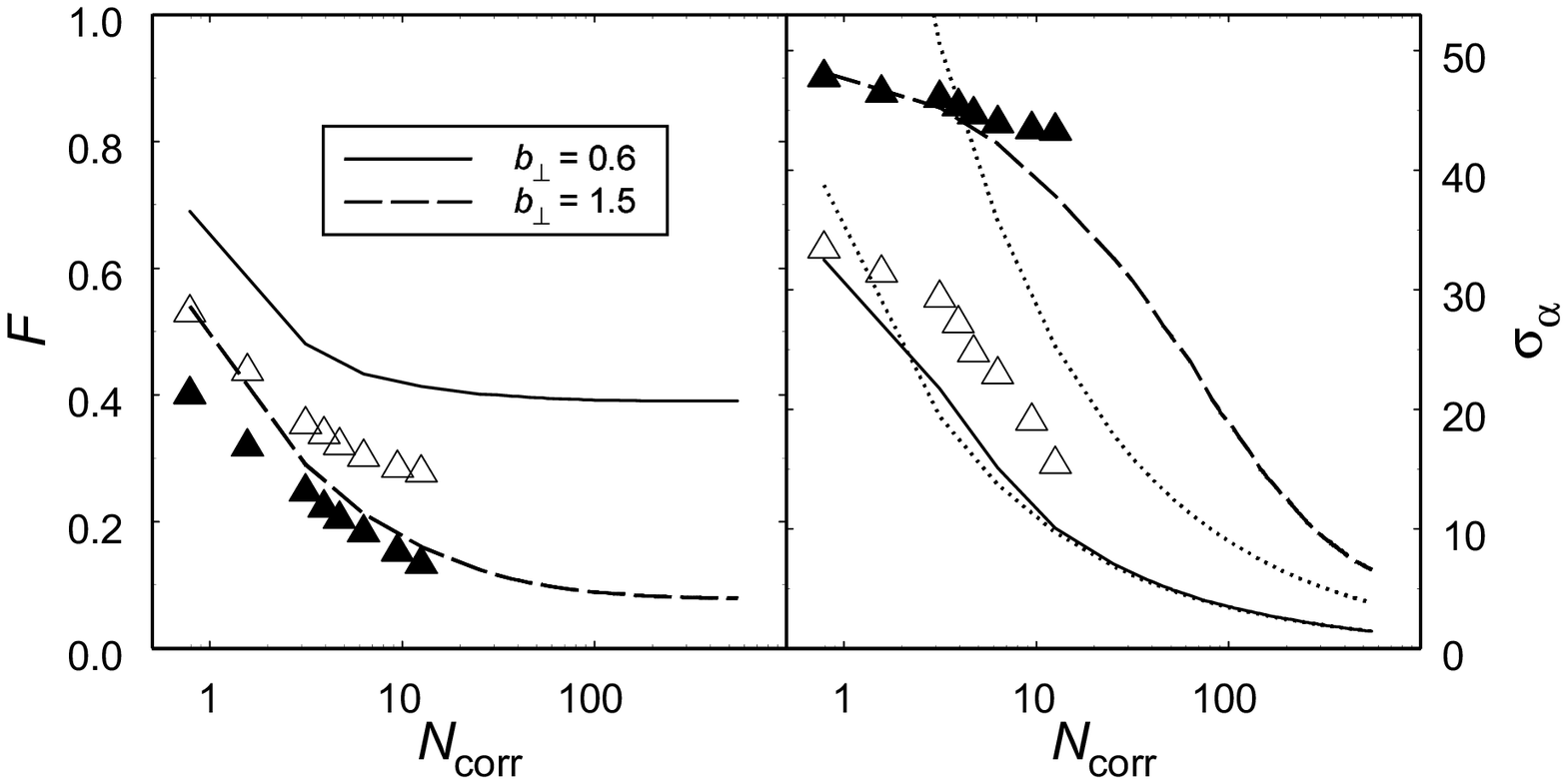}
\caption{Polarization reduction factor $F$ due to turbulence and the dispersion $%
\protect\sigma _{\protect\alpha }$ in the position angle vs the number of
correlation lengths $N_{\mathrm{corr}}$ through a cloud. Results from
the actual MHD computations are denoted by filled ($b_{\bot}=1.5$) and open ($%
b_{\bot}=0.6$) triangles. The dispersion of position angles obtained with the expression of
\protect{\cite{mg91}} is indicated by the dotted lines.}
\label{main}
\end{figure}

\begin{figure}[t]
\epsscale{0.4} \plotone{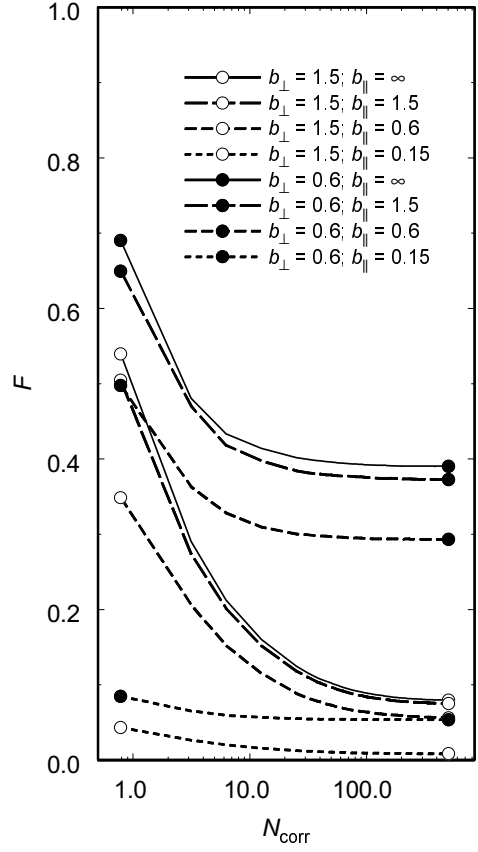}
\caption{Polarization reduction factor $F$ due to turbulence and with a non-zero LOS
magnetic field. Weak and strong field cases are indicated by open and filled circles,
respectively. Solid lines ($b_{\Vert }=\infty$) are computed when the uniform
magnetic field component along the LOS is zero.}
\label{los}
\end{figure}

\begin{figure}[t]
\epsscale{0.5} \plotone{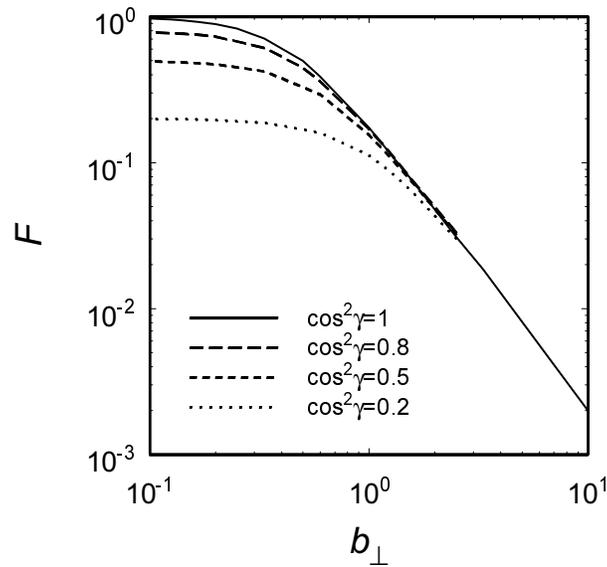}
\caption{Asymptotic values of $F$ ($\equiv F_{\infty})$ versus $b_{\bot}$ for
representative values of $\cos ^{2}\protect\gamma$.}
\label{f_vs_b}
\end{figure}

\begin{figure}[t]
\epsscale{1.} \plotone{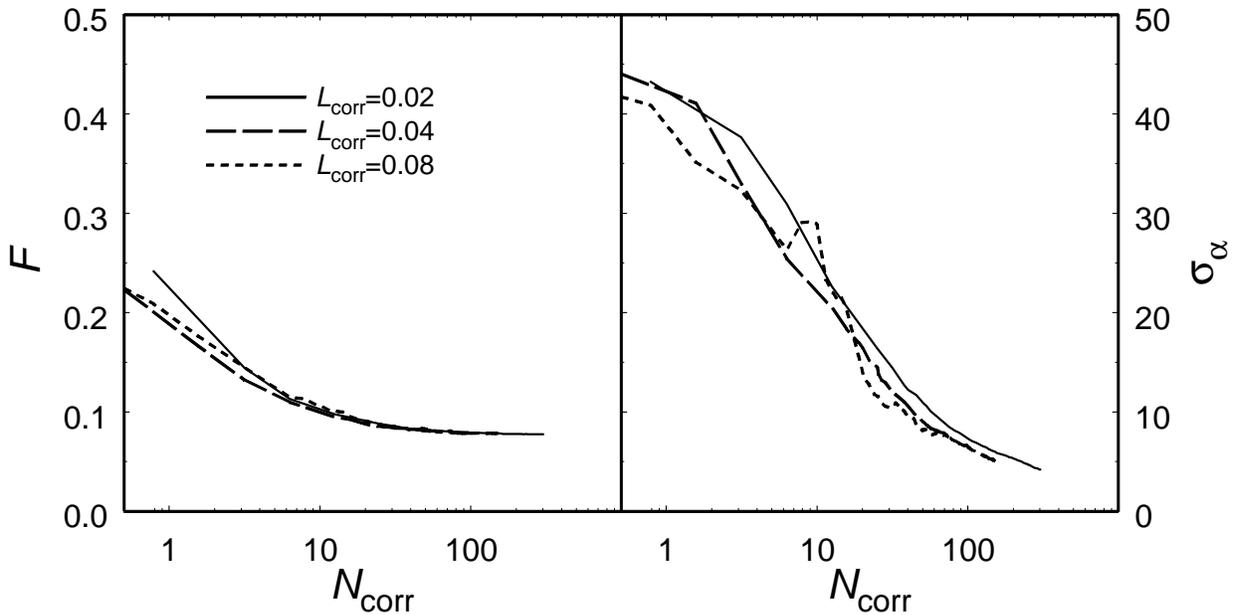}
\caption{Polarization reduction factor $F$ due to turbulence and 
the dispersion $\sigma_\alpha$ in the position angles vs the number of 
correlation lengths $N_{\mathrm{corr}}$ through the cloud for various
values of the correlation length and $b_{\perp}=1.5$. The ratio of the beam size to $L_{\mathrm{corr}}$
is the same for all three sets of curves.}
\label{beamrat}
\end{figure}

In the context of the polarization hole effect,
Figure~\ref{main} suggests that the decrease in the percentage
polarization can be a result of the larger number of correlation lengths
that a ray traverses in passing through a dense core versus the number it 
traverses in the 
surrounding medium. For this, $N_\mathrm{corr}$ in the surrounding medium
must be small because the polarization reduction factor $F$ reaches its asymptotic
value quite fast. For the moderately strong uniform field ($b_{\bot}=0.6$), $F$ is 
essentially $%
F_\infty$ when $N_\mathrm{corr}$ is approximately a few. In the
case of the weak uniform field ($b_{\bot}=1.5$) that we favored in \cite{abs}, a
factor of a few decrease in $F$ is achieved between $N_\mathrm{corr}\sim1$
and 30. Over the same range of $N_\mathrm{corr}$, $\sigma_\alpha$ decreases
from 40$^\circ$ to about 20$^\circ$.

\begin{figure}[t]
\plottwo{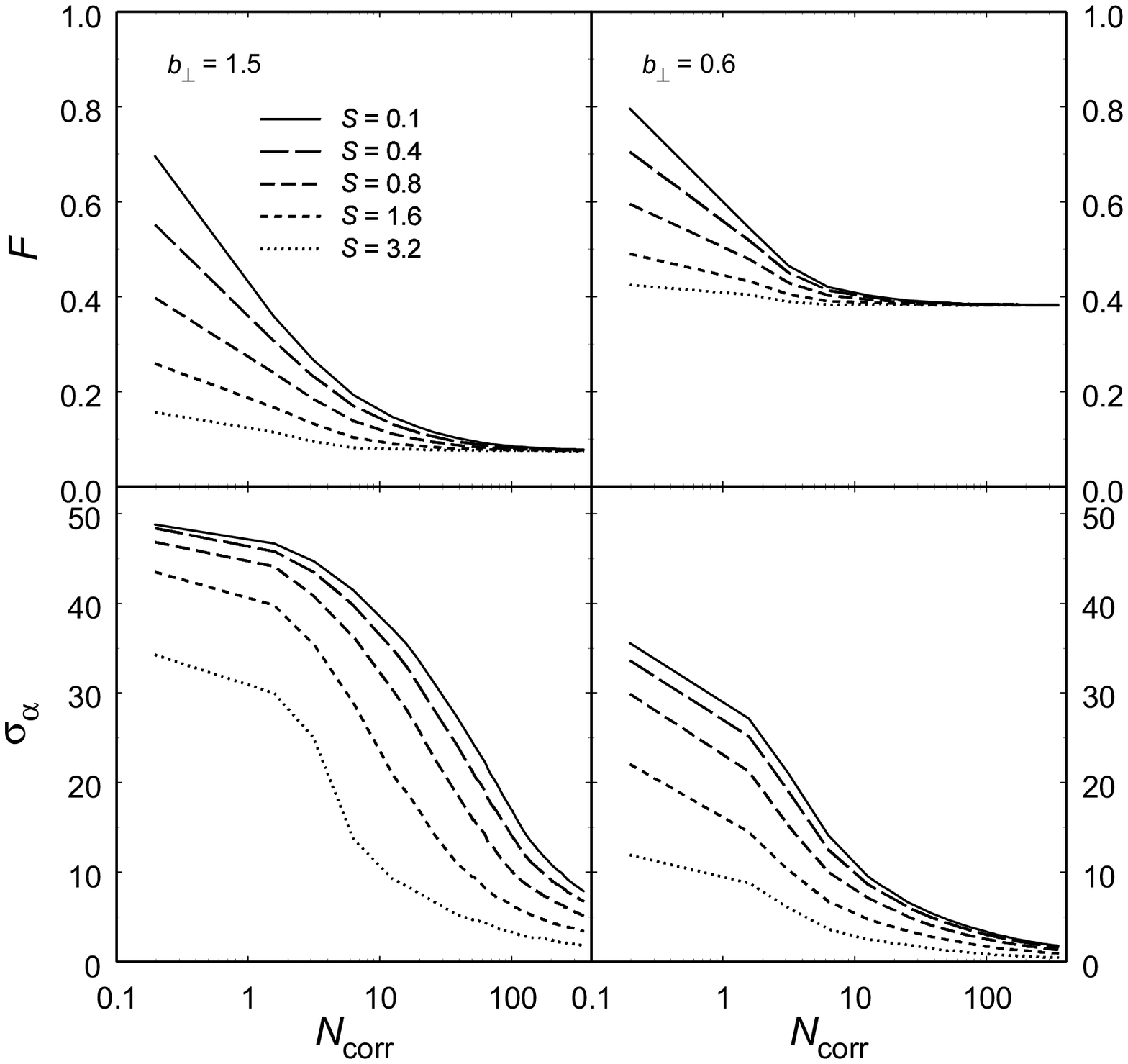}{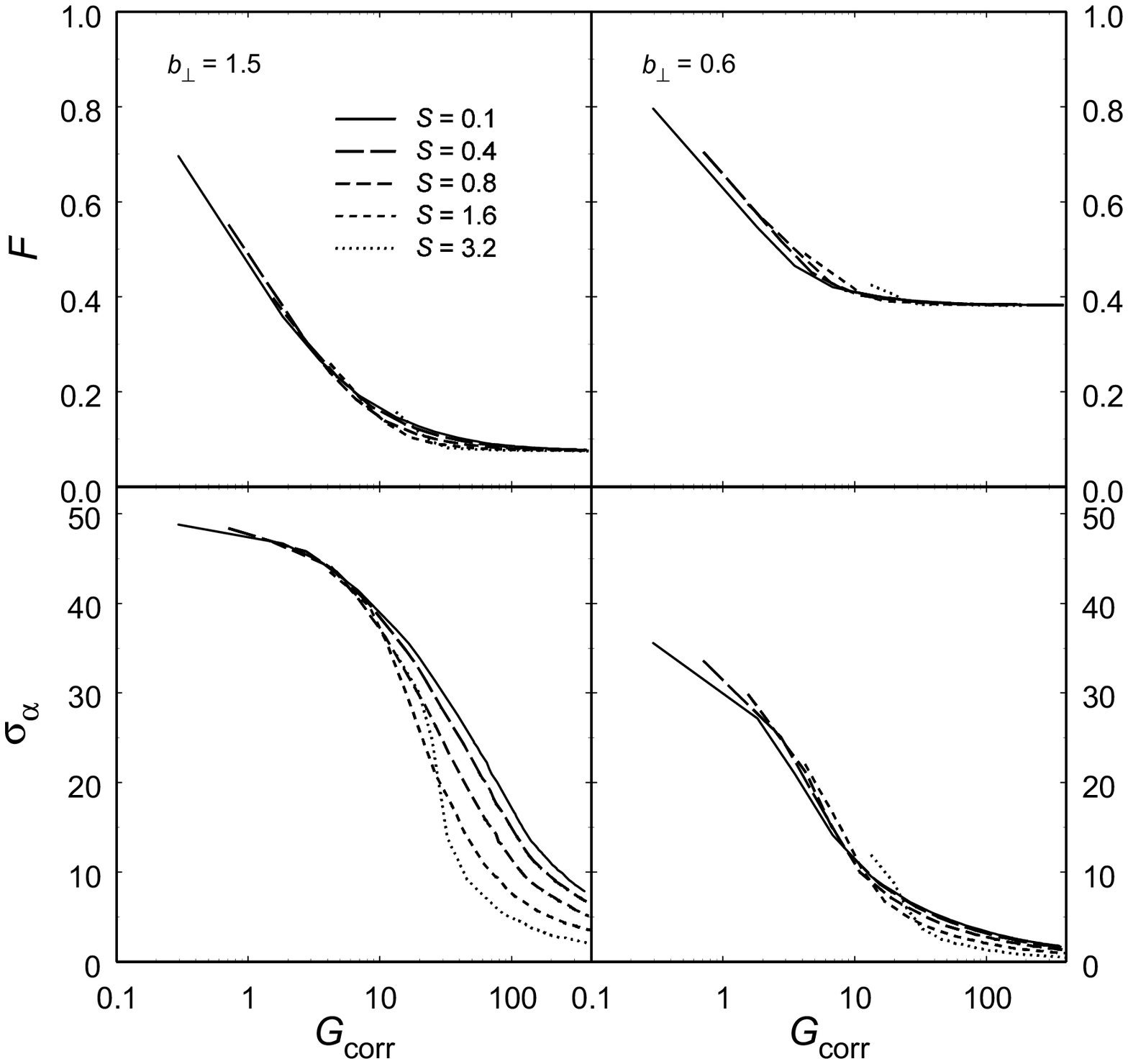}
\caption{Left panel--- $F$ and $\protect\sigma_{%
\protect\alpha}$ vs the number of correlation lengths $N_{\mathrm{corr}}$ 
through a cloud for
different magnetic field ratios and beam sizes. Right panel---Same
quantities as in left panel, but shown as functions of a generalized number
of correlation lengths $G_{\mathrm{corr}}$.}
\label{beam}
\end{figure}

These values for $N_\mathrm{corr}$ and $\sigma_\alpha$ are larger
than the values that are indicated from the observations. The need to have $N_\mathrm{%
corr}$ near 30 is relaxed if we take into account a finite beam size and
allow the correlation length $L_\mathrm{corr}$ to be smaller in the core than
on the periphery.
Let us assume that on the periphery of a dense core,  a ray traverses 
$N_\mathrm{corr}\sim1-3$ and that the beam size is smaller than $L_\mathrm{%
corr}$. In this case $F$ is about $0.4-0.5$ (the solid or long-dashed curve on the top
left panel of Figure~\ref{beam}). If toward the center of the dense core $L_\mathrm{corr}$
becomes smaller than the beam size, then $F\sim0.1$ toward the center (the dotted curve on 
the top left panel of
Figure~\ref{beam}). With this beam size and with $N_\mathrm{corr}\sim 10$, 
$\sigma_\alpha$ becomes less than 10$^\circ$. If the same region is observed with
a smaller beam (e.g., with an interferometer), then the dispersion in
position angles would increase (because $G_\mathrm{corr}$ becomes smaller).

The apparent discrepancy in this scenario of a large dispersion in position angles at the 
periphery of the core ($\sim40^\circ$)
can be partially circumvented by assuming that the strength of the uniform
magnetic field is higher in the vicinity of the core than in the core
itself. When $b_{\bot}=0.6$ in the periphery region, $\sigma_\alpha$ can
be as small as 20$^\circ$ (solid curve on the lower right panel of Figure~%
\ref{beam} at $N_\mathrm{corr}=3$). It is even smaller at lower $b_{\bot}$.

The requirement that a ray traverse only a few correlation lengths in a region
surrounding the dense core where the polarization hole is observed may also seem
to conflict with our conclusion in \cite{abs} that a typical dark
cloud spans no fewer than 10~correlation lengths. It must be noted
that in the B1 cloud, where polarimetry observations of both starlight and 
thermal dust emission are available, they indicate different directions 
for the magnetic field
\citep{MW2002}. It is quite possible that the polarized light of background
stars and the polarized dust emission, even when they are observed
close to one another, trace different objects or parts of the same object
that differ in physical conditions and/or scale. The two types of observations
may thus not be directly
related to one another. In addition,  observations of starlight polarization
are mostly
used to study quiescent dark clouds, while many (though not all) objects in
which the polarization of the dust emission is observed already
contain Class~0 protostars with outflows.

\begin{figure}[t]
\plotone{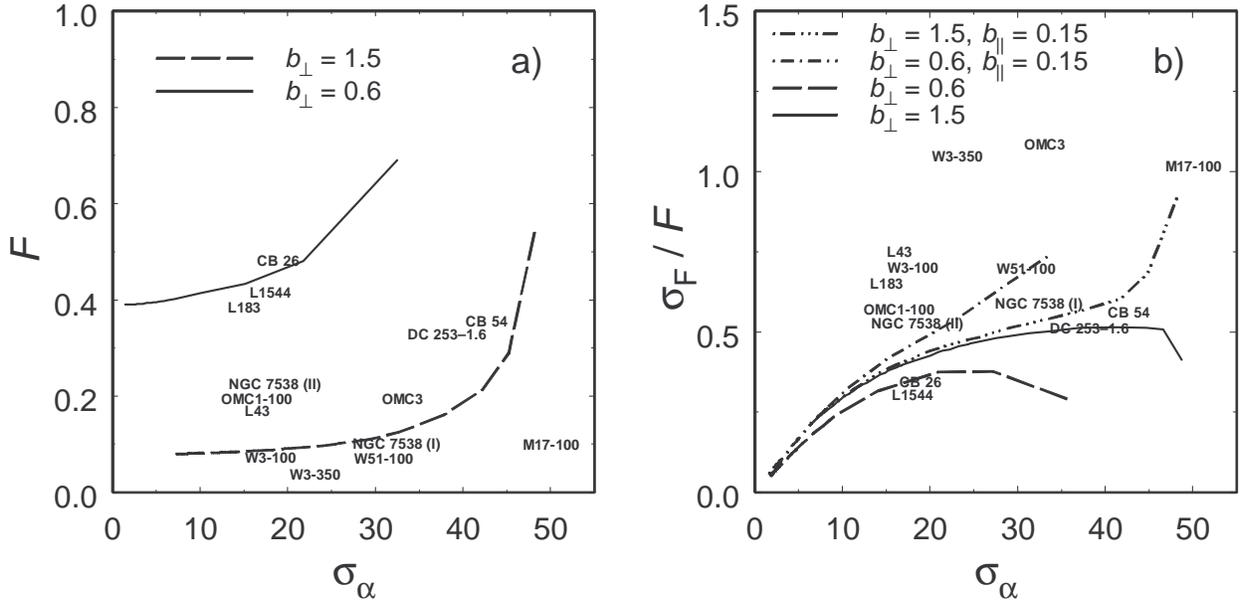}
\caption{Computed polarization reduction factor $F$ due to turbulence (left panel)
and the relative polarization error $\sigma_F/F$ (right panel) as a function of
$\sigma_{\alpha}$.
Observational data also are shown for comparison. The name of the object is
centered on the corresponding data point. The number 100 or 350 next to the name of the object
indicates the wavelength in $\protect\mu $m. If no number is specified, the 
measurement is at 850~$\protect\mu $m. Some cloud labels have been shifted slightly for
clarity}
\label{f_vs_sigmaa}
\end{figure}

\begin{figure}[t]
\epsscale{0.7} \plotone{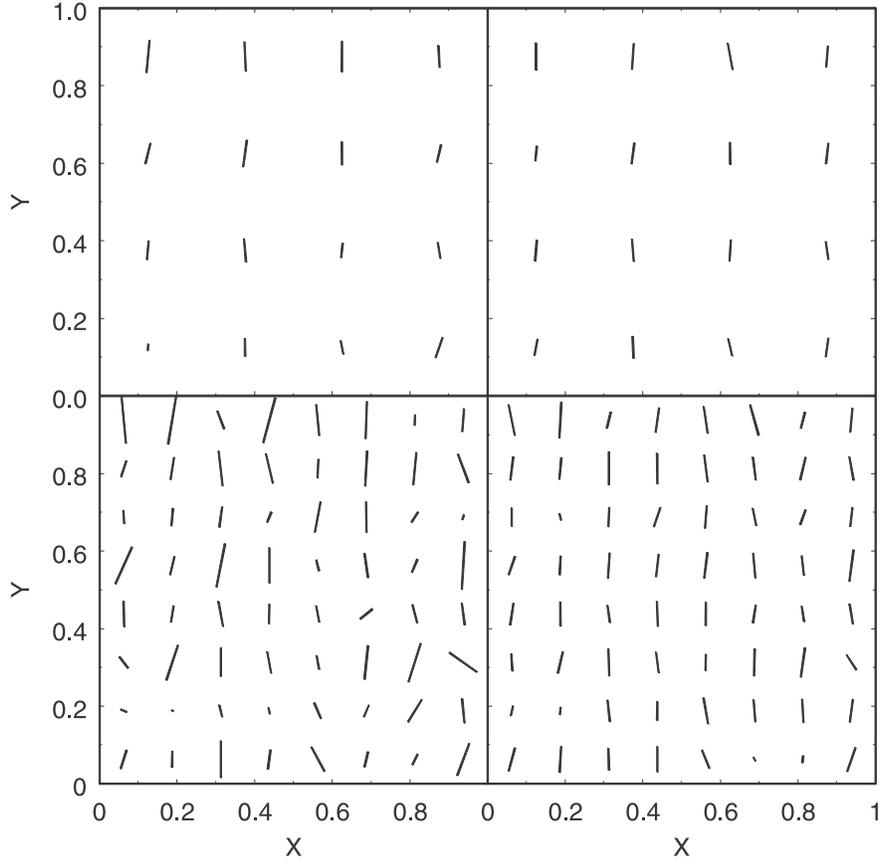}
\caption{Computed maps of polarization vectors with the weak, uniform magnetic field
for two values of $N_{\rm corr}$ (left column, $N_{\rm corr}=12$; right column, 
$N_{\rm corr}=36$) and for two values of $S$ (top row, $S=3.2$; bottom
row, $S=1.6$). Coordinates are given as a fraction of the distance along the face of 
the computational cubes. The longest vector corresponds to $F=0.18$.
The uniform field is in the $x$ direction. Position angle dispersions are $8^{\circ}$
(left) and $7^{\circ}$ (right) for the top row, and $21^{\circ}$ (left)
and $11^{\circ}$ (right) for the bottom row.}
\label{map}
\end{figure}

\begin{figure}[ht]
\begin{center}
\includegraphics[width=0.6\hsize]{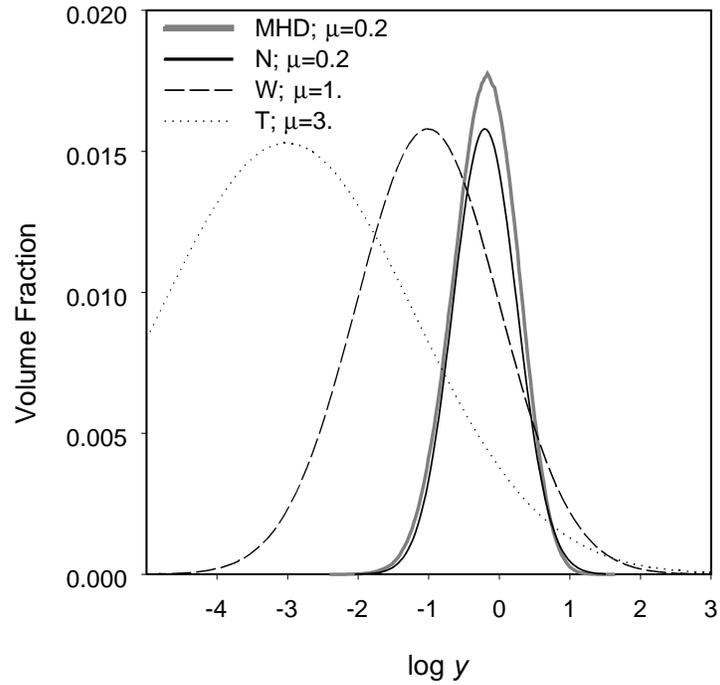}
\end{center}
\caption{Density distributions that are utilized in the computations. 
The distribution `N'
(`narrow') is chosen to be as close as possible to the results of
Stone et al. (1998), which are indicated by the thick gray line. Distribution `W' (`wide')
is intended to represent results of other numerical simulations that
lead to a wider density PDF. The distribution used to study
thermalization effects is labelled with `T'.}
\label{dustdis}
\end{figure}

\begin{figure}[t]
\begin{center}
\includegraphics[width=0.6\hsize]{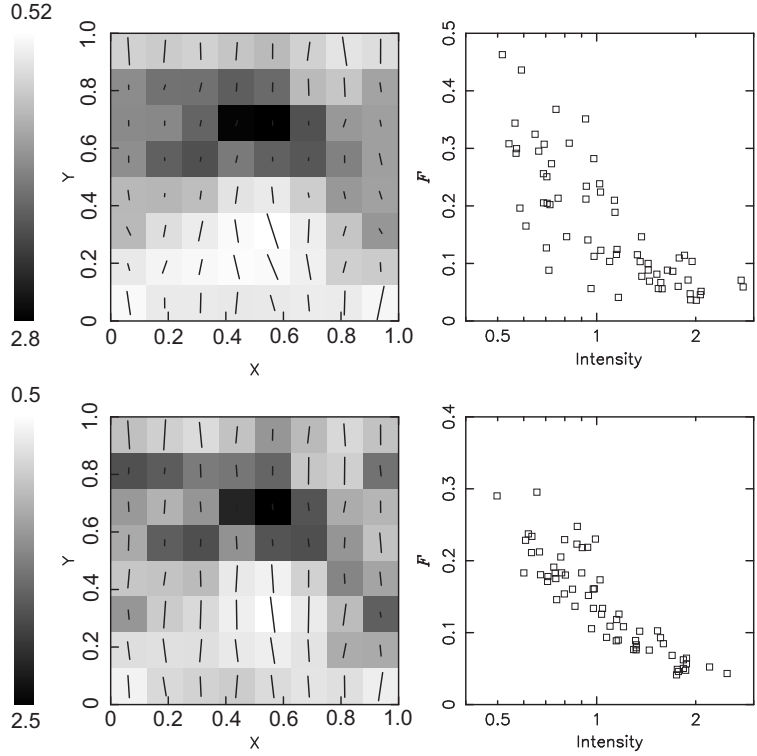}
\end{center}
\caption{Map of polarization vectors and $F$ vs $I$ diagram for the `N' distribution
and the strong uniform magnetic field. Upper panel, $N_\mathrm{corr}=5$, lower panel, $N_%
\mathrm{corr}=12$. The longest vector corresponds to the highest $F$. The map is
obtained by averaging over $16\times16$ emerging rays. The grayscale on left panel and the scale of 
the horizontal axis on right panel is the intensity of dust emission in arbitrary units.
The uniform magnetic field is in the $x$ direction.}
\label{narrowmed}
\end{figure}

\begin{figure}[t]
\begin{center}
\includegraphics[width=0.6\hsize]{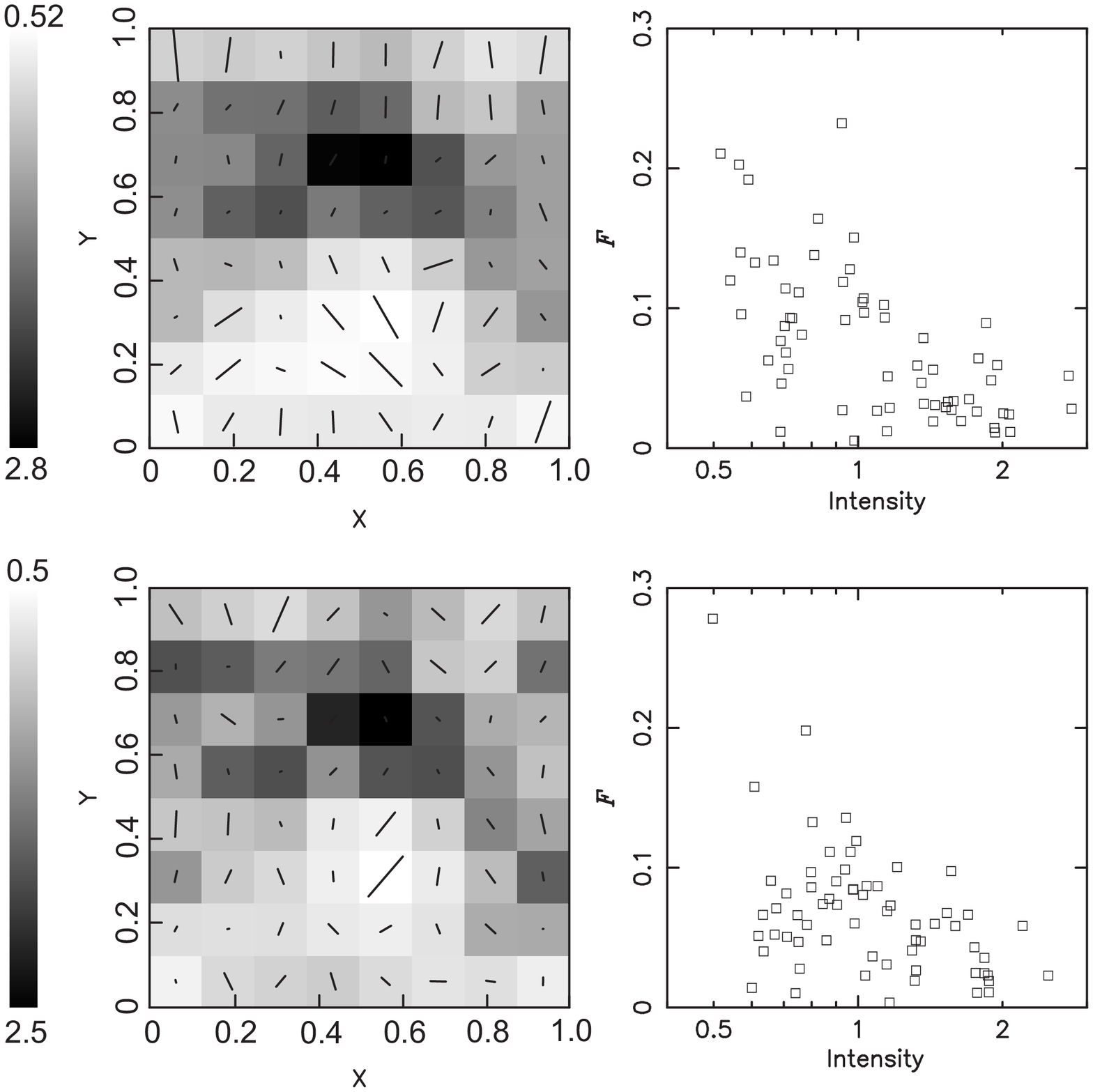}
\end{center}
\caption{Same as in Figure~\ref{narrowmed}, but for the weak uniform magnetic field.}
\label{narrowweak}
\end{figure}

\begin{figure}[t]
\begin{center}
\includegraphics[width=0.6\hsize]{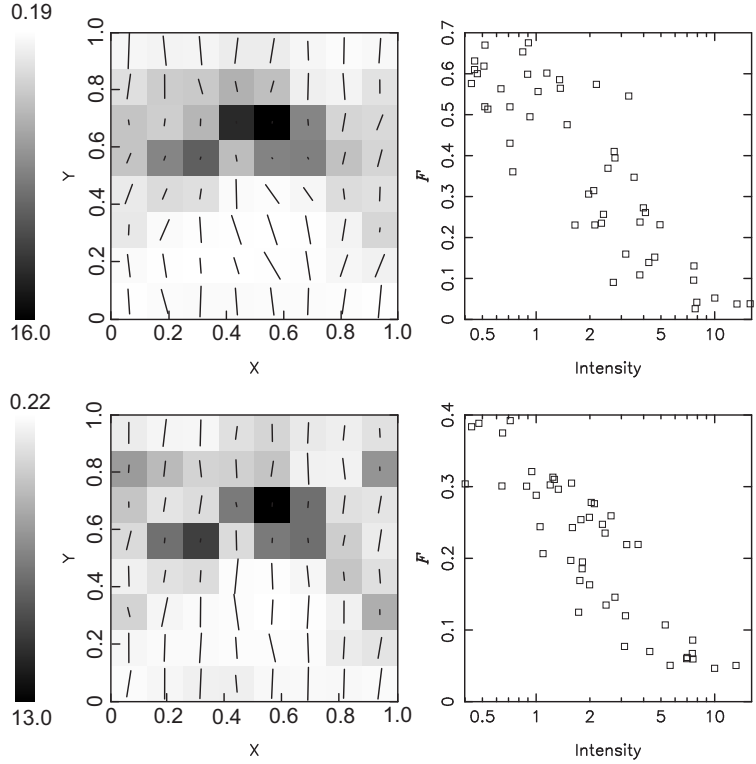}
\end{center}
\caption{Map of polarization vectors and $F$ vs $I$ diagram for the `W' distribution
and the strong uniform magnetic field. Upper panel, $N_\mathrm{corr}=5$, lower panel, $N_%
\mathrm{corr}=12$. Intensity is in arbitrary units. The longest vector
corresponds to the highest $F$. The map is
obtained by averaging over $16\times16$ emerging rays. The grayscale on left panel and the scale of 
the horizontal axis on right panel is the intensity of dust emission in arbitrary units. The 
uniform magnetic field is in the $x$ direction.}
\label{widemed}
\end{figure}

\begin{figure}[t]
\begin{center}
\includegraphics[width=0.6\hsize]{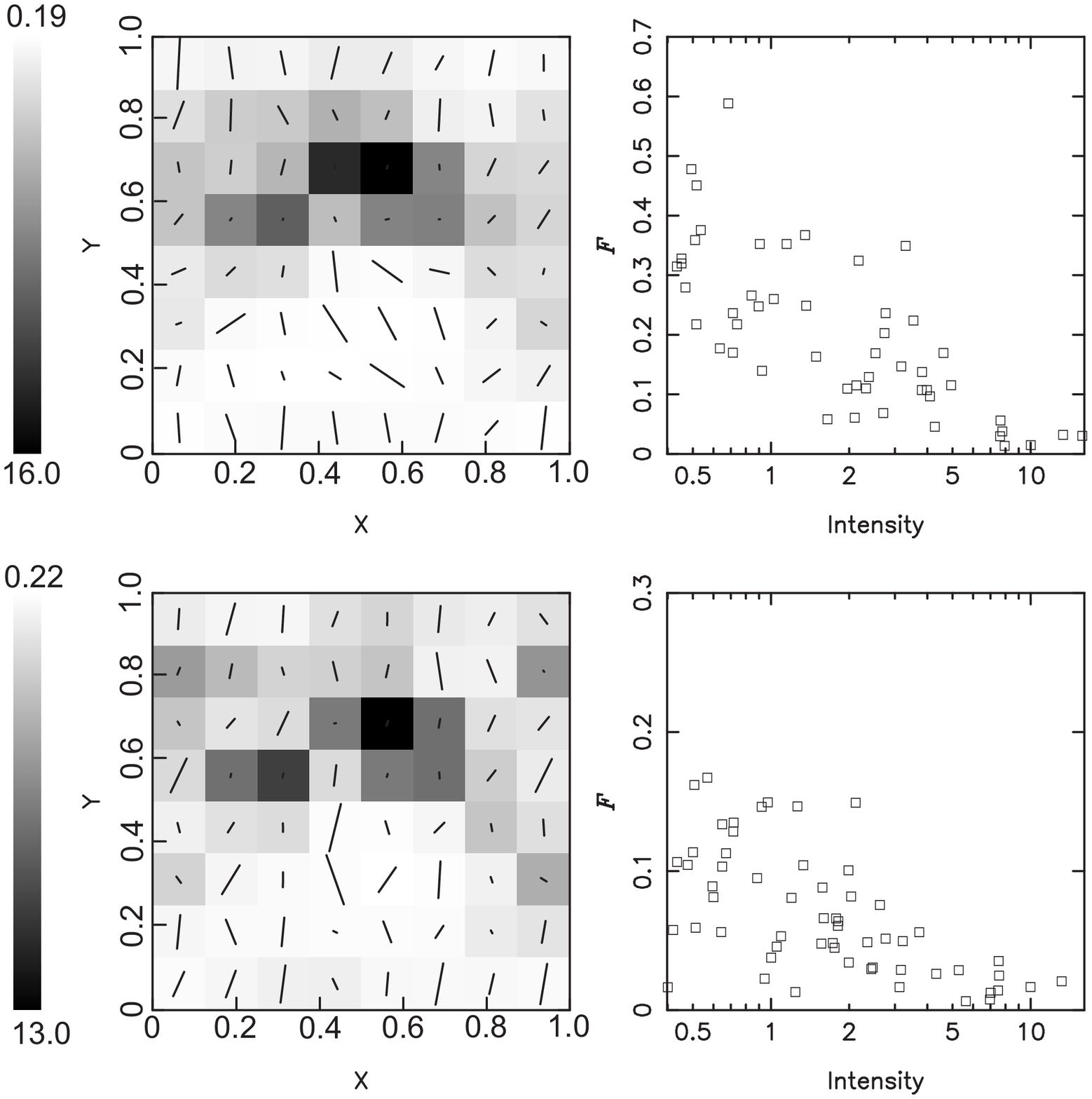}
\end{center}
\caption{Same as in Figure~\ref{widemed}, but for the weak uniform magnetic field.}
\label{wideweak}
\end{figure}

\begin{figure}[t]
\begin{center}
\includegraphics[width=0.6\hsize]{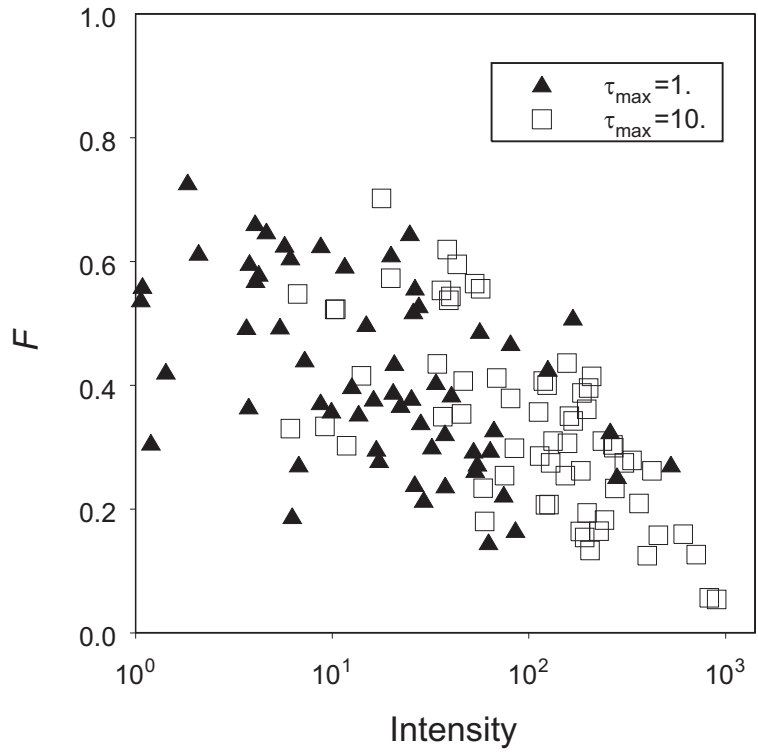}
\end{center}
\caption{Polarization reduction due to effects of thermalization.}
\label{thermal}
\end{figure}

\begin{figure}[t]
\begin{center}
\includegraphics[width=0.6\hsize]{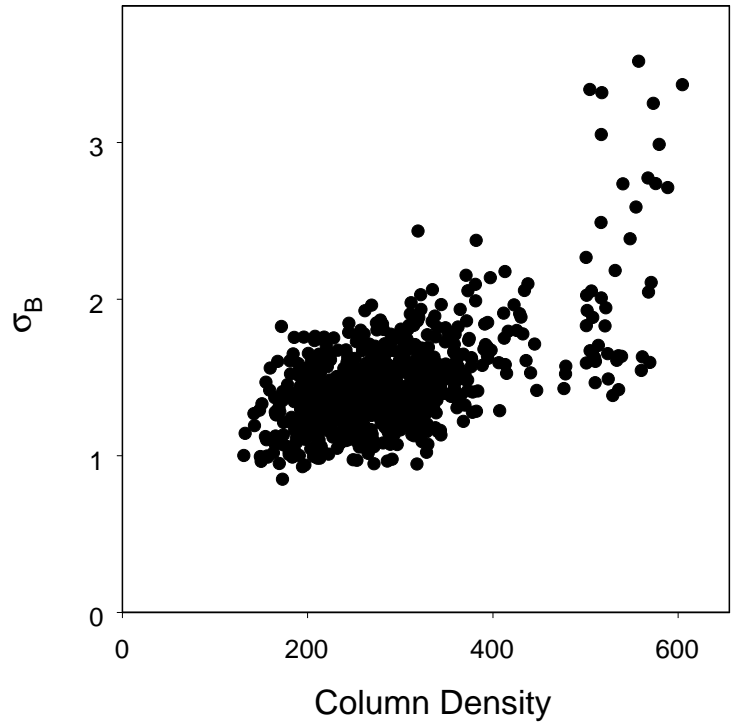}
\end{center}
\caption{The rms magnetic field versus column density in the MHD
simulations by Stone et al. (1998), in arbitrary units}
\label{mfcolden}
\end{figure}

\begin{figure}[t]
\begin{center}
\includegraphics[width=0.6\hsize]{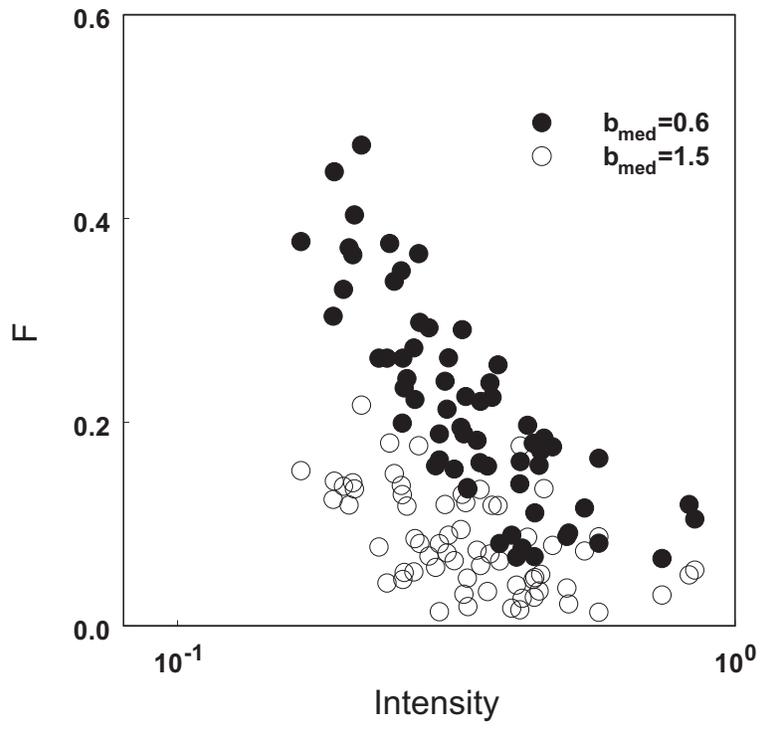}
\end{center}
\caption{Polarization reduction factor $F$ as a function of intensity for
variable $b_{\bot}$.}
\label{bd}
\end{figure}

Our calculations demonstrate quantitatively how the magnitude of the correlation length
(or at least how the correlation length compares with the size of a cloud and 
the size of the telescope beam)
is the key factor for the influence of turbulence on the observations of the linear 
polarization. Unfortunately, any suggestions based on our 
calculations that we can offer for extracting 
the correlation length from observational data are hampered
by the lack of information about $R$ and about the source geometry.
With these limitations in mind, some inferences about $N_{\rm corr}$ can be made
based on Figures~\ref{beam} and~\ref{f_vs_sigmaa} and an application of equation 
~(\ref{gcorr}). Assuming that the
number of correlation lengths is the same along the line of sight and perpendicular to it,
we may replace $S$ in equation~(\ref{gcorr}) with $N_{\rm corr}/D$, where $D$ is the number of beams
across an object. Equation~(\ref{gcorr}) then can be solved for $N_{\rm corr}$ in 
terms of $G_{\rm corr}$.

Let us consider two examples. The location of the L1544 core in left panel of Figure~\ref{f_vs_sigmaa}
implies that the uniform magnetic field is strong in this object. The polarization position angle
dispersion $\sigma_\alpha$ then corresponds to $G_{\rm corr}\sim10$ (see right panel of Figure~\ref{beam}).
For these observations, $D\sim8$ and we obtain $N_{\rm corr}\sim5$. Assuming
the generally accepted distance of 140 pc and an angular size of about $110^{\prime\prime}$
\citep{firstmap} gives
$L_{\rm corr}\sim0.02$~pc for this object. The L43 core, on the other hand, in left panel
of Figure~\ref{f_vs_sigmaa} is located close to the curve corresponding to the weak uniform field.
The small anglular dispersion implies $G_{\rm corr}\sim50$ according to Figure 5 (right panel), 
which for the same $D$ gives $N_{\rm corr}\approx20$.
At a distance of 170 pc and an angular size of 
$\sim100^{\prime\prime}$ \citep{firstmap}, $L_{\rm corr}\sim0.004$~pc.
Determinations of the correlation lengths by independent methods would be valuable.
Unfortunately, the tendency for velocity shifts in clouds to be caused by large-scale
velocity gradients as well as by turbulence, for the presence of inhomogeneities 
in cloud properties, etc. obscure the detailed effects of turbulence .

This work was supported in part by NSF Grant AST 9988104. DW acknowledges support
from the President of the RF grant MK-348.2003.02.

\clearpage

\end{document}